\documentclass[12pt]{article}
\usepackage{epsfig}
\usepackage{here}
\usepackage{a4}
\usepackage{amssymb}

\begin{document}

\def\Journal#1#2#3#4{{#1} {\bf #2}, #3 (#4)}

\def\etal{{\it et\ al.}}
\def\NCA{\em Nuovo Cim.}
\def\NIM{\em Nucl. Instrum. Methods}
\def\NIMA{{\em Nucl. Instrum. Methods} A}
\def\NPB{{\em Nucl. Phys.} B}
\def\PLB{{\em Phys. Lett.}  B}
\def\PRL{\em Phys. Rev. Lett.}
\def\PRC{{\em Phys. Rev.} C}
\def\PRD{{\em Phys. Rev.} D}
\def\ZPC{{\em Z. Phys.} C}
\def\ASP{{\em Astrop. Phys.}}
\def\JETP{{\em JETP Lett.\ }}

\def\numunue{\nu_\mu\rightarrow\nu_e}
\def\numunutau{\nu_\mu\rightarrow\nu_\tau}
\def\nuebar{\bar\nu_e}
\def\nue{\nu_e}
\def\nutau{\nu_\tau}
\def\numubar{\bar\nu_\mu}
\def\numu{\nu_\mu}
\def\ra{\rightarrow}
\def\numubarnuebar{\bar\nu_\mu\rightarrow\bar\nu_e}
\def\nuebarnumubar{\bar\nu_e\rightarrow\bar\nu_\mu}
\def\osc{\rightsquigarrow}

\def\inteni{{\cal I}_{pot}}
\def\fmerit{{\cal F}}

\thispagestyle{empty}
\begin{flushright}
{\tt ICARUS/TM-2002/09}\\ 
July 5, 2002
\end{flushright}
\vspace*{0.5cm}
\begin{center}
{\Large{\bf A low energy optimization of the CERN-NGS
neutrino beam for a $\theta_{13}$ driven neutrino oscillation
search} }\\
\vspace{.5cm}
A. Rubbia
and P. Sala\footnote{On leave of absence from INFN Milano.}

\vspace*{0.3cm}
Institut f\"{u}r Teilchenphysik, ETHZ, CH-8093 Z\"{u}rich,
Switzerland\\
\end{center}
\vspace{0.3cm}
\begin{abstract}
\noindent
The possibility to improve the CERN to Gran Sasso neutrino beam 
performances for  $\theta_{13}$ searches is investigated. 
We show that by an appropriate optimization of the target
and focusing optics of the present CNGS design,
we can increase the flux of low energy neutrinos by
about a factor 5 compared to the current $\tau$ optimized focalisation.
With the ICARUS 2.35 kton detector
at LNGS and in case of negative result, this would allow to improve the
limit to $\sin^22\theta_{13}$ by an order of magnitude
better than the current limit of CHOOZ at $\Delta m^2\approx 3\times 10^{-3}\ \rm eV^2$
within 5 years of nominal CNGS running. This is by far
the most sensitive setup of the currently approved long-baseline experiments
and is competitive with the proposed JHF superbeam.
\end{abstract}

\pagestyle{plain} 
\setcounter{page}{1}
\setcounter{footnote}{0}

\section{Introduction}
The firmly established disappearance of muon neutrinos of cosmic ray 
origin~\cite{kamio,Fukuda:1998mi} strongly points 
toward the existence of neutrino
oscillations.

The approved first generation long baseline (LBL) experiments ---
K2K~\cite{k2k}, MINOS~\cite{minos}, 
ICARUS~\cite{icarus} and OPERA~\cite{opera} --- will search for
a conclusive and unambiguous signature of the oscillation
mechanism.
They will provide the first precise
measurements of the parameters governing the main
muon disappearance mechanism. 
In particular, the CERN-NGS beam\cite{CNGS,Addendum}, 
specifically optimized for tau appearance,
aims to directly confirm the hints for neutrino flavor oscillation.

The physics program of ICARUS will start 
with the installation
of the 600 ton prototype at the Gran Sasso Laboratory (LNGS) and will allow the
observation of atmospheric neutrinos, the detection of solar and supernovae neutrinos
and the search for proton decay. An extension of the mass
of argon is foreseen with the goal of reaching a total mass of about 
3000 tons. 

In addition to the dominant $\numu\ra\nutau$ oscillation, it is possible that a sub-leading
transition involving electron-neutrinos occur as well.
In the ``standard interpretation'' of the 3-neutrino mixing, the $\numu \ra \nue$ oscillations at the 
$\Delta m^2 \approx 2.5\times 10^{-3}\rm\ eV^2$ indicated
by atmospheric neutrinos is driven by the so-called $\theta_{13}$ angle.
Indeed, given
the flavor eigenstates
$\nu_\alpha(\alpha= e,\mu,\tau)$ related to the mass eigenstates
$\nu'_i(i=1,2,3)$ where
$\nu_\alpha=U_{\alpha i}\nu'_i$, the mixing matrix $U$ is parameterized as:
\begin{equation}
U(\theta_{12},\theta_{13},\theta_{23},\delta)=\left(
\begin{tabular}{ccc}
$c_{12}c_{13}$      & $s_{12}c_{13}$   &  $s_{13}e^{-i\delta}$ \\
$-s_{12}c_{23}-c_{12}s_{13}s_{23}e^{i\delta}$ &
$c_{12}c_{23}-s_{12}s_{13}s_{23}e^{i\delta}$ & $c_{13}s_{23}$ \\
$s_{12}s_{23}-c_{12}s_{13}c_{23}e^{i\delta}$ &
$-c_{12}s_{23}-s_{12}s_{13}c_{23}e^{i\delta}$ & $c_{13}c_{23}$ 
\end{tabular}\right)
\end{equation}
with $s_{ij}=\sin\theta_{ij}$ and $c_{ij}=\cos\theta_{ij}$.

The best sensitivity for this oscillation is 
expected for ICARUS at the CERN-NGS.
Limited by the CNGS beam statistics at low energy, this search
should allow to improve by roughly a factor 5 (see Ref.\cite{icarus}) 
the CHOOZ\cite{Chooz}
limit on the $\theta_{13}$ angle 
for $\Delta m^2 \approx 3\times 10^{-3}\rm\ eV^2$.
Beyond this program, 
new methods will be required in order to improve significantly
the sensitivity. 

At present, the only well established proposal in this direction 
is the JHF-Kamioka project\cite{JHF}. 
In its first phase, 5 years of operation with
the Super-K detector, it aims to a factor 20 improvement
over the  CHOOZ limit.

In this paper, 
we  investigate the possibility to perform $\theta_{13}$ searches
with a CERN to Gran Sasso beam  
optimized for Low Energy (L.E.) neutrino production. This is meant to be 
{\it alternative} to the CNGS $\tau$ program, 
in order to be competitive with JHF.

FLUKA\cite{FLUKA1,FLUKA2} Monte Carlo 
simulations are employed to calculate neutrino yields with  specific
target and focusing configuration.   

Simulated neutrino fluxes are employed to derive exclusion plots in the 
$\Delta m^2 $ vs $\sin^2(2\theta_{13})$ plane for the ICARUS detector.

\section{The neutrino energy range}
In order to maximize the probability of an oscillation, 
we must choose the energy of the neutrino $E_{max}$ and
the baseline $L$ such that 
\begin{equation}
1.27\frac{L(km)}{E_{max}(GeV)}\Delta m^2(eV^2)\simeq \frac{\pi}{2}
\end{equation}
However, in order to {\it observe
the oscillation}, we must at least see the maximum preceded by a minimum,
 given by 
\begin{equation}
1.27\frac{L(km)}{E_{min}(GeV)}\Delta m^2(eV^2)\simeq \pi
\end{equation}
At the CERN-LNGS baseline (730 Km), with  $\Delta m^2$ in the range  
indicated by Super Kamiokande 
($ 1\times 10^{-3}<\Delta m^2 < 4\times 10^{-3}\rm\ eV^2$), the maxima and
minima of the oscillation lie between 0.3 GeV and 2.4 GeV 
(See Table~\ref{tab:enenu}). 

\begin{table}[tb]
\begin{center}
\begin{tabular}{|c||c|c||c|c||c|c||c|c|}
\hline
&\multicolumn{8}{c|}{$\Delta m ^2 $(eV $^2$)}\\
\hline
L&\multicolumn{2}{c|}{$1\times 10^{-3}$}&\multicolumn{2}{c|}{$2\times 10^{-3}$}&
\multicolumn{2}{c|}{$3\times 10^{-3}$}&\multicolumn{2}{c|}{$4\times 10^{-3}$}\\
(km)&E$_{max}$  & E$_{min}$  &E$_{max}$  & E$_{min}$  
&E$_{max}$  & E$_{min}$  &E$_{max}$  & E$_{min}$  \\
&MeV&MeV&MeV&MeV&MeV&MeV&MeV&MeV\\
\hline
 730  &   590  &   295  &  1180  &   590  &
  1771  &   885  &  2361  &  1180\\
\hline
\end{tabular}
\end{center}
\caption{Neutrino energies $E_{max}$ and $E_{min}$ (see text for definition) 
corresponding to
the maximum and minimum of the $\nu_\mu\rightarrow\nu_x$ oscillation
for the CERN-LNGS baseline and the $\Delta m^2$ range indicated by
Superkamiokande.}
\label{tab:enenu}
\end{table}


\section{The present CERN-LNGS configuration}
The present CNGS design\cite{CNGS} is optimized for 
$\nu_\tau$ appearance, thus for a 
relatively high-energy neutrino beam. The 400 GeV/c SPS  beam will
nominally deliver $4.5\times 10^{19}$ protons per year on  a
graphite target, made of spaced rods to reduce the re-interaction rate within
the target.
The two magnetic horns (horn and reflector) are tuned to focus 35 and 50
GeV/c mesons, with an acceptance of the order of 30~mrad. The present
shielding and collimator openings would not allow more than 100~mrad even
in perfect focusing.   
The decay tunnel length is 1 km, and the baseline for neutrino oscillation
is 732 km. 
 
\section{The CNGS L.E. option}
To improve  particle yield at low energies, we re-designed the focusing
system, and changed the target dimensions and shortened the effective
decay tunnel length. The main
differences wit the present ($\tau$) design are summarized in
Table~\ref{tab:parameters}.

\begin{table}[htb]
\begin{center}
\begin{tabular}{|l|c|c|}
\hline
&CNGS $\tau$ &CNGS L.E.\\
{\bf Target}&&\\
Material & Carbon&Carbon\\
Total target length & 2 m &1 m \\
Number of rods & 13 & {\bf 1} \\
Rod spacing& first 8 with 9 cm dist.& none\\
Diameter of rods& first 2 5 mm, then 4 mm& {\bf 4mm} \\
&&\\
{\bf Horn }&&\\
Distance beginning of target-horn entrance &320 cm &{\bf 25 cm}\\
Length &6.65 m & {\bf 4 m}\\
Outer conductor radius& 35.8 cm&80 cm $^\dagger$\\
Inner conductor max. radius&6.71 cm&11.06 cm\\
Inner conductor min. radius&1.2 cm&0.2 cm\\
Current&150kA&300kA\\
&&\\
{\bf Reflector }&&\\
Distance beginning of target-reflector entrance &43.4 m &{\bf 6.25 m}\\
Length &6.65 m & {\bf 4 m}\\
Outer conductor radius& 55.8 cm&90 cm $^\dagger$\\
Inner conductor max. radius&28 cm&23.6 cm\\
Inner conductor min. radius&7cm&5 cm\\
Current&180kA&150kA\\
&&\\
{\bf Decay tunnel }&&\\
Distance beginning of target-tunnel entrance &100 m &50 m\\
    Length &992 m &350 m\\
 Radius & 122 cm &350 cm $^\dagger$ \\
\hline
\end{tabular}
\end{center}
\caption{Parameter list for the present CNGS design
and the ``new'' beam  for low energy $\nu$ . For the parameters 
flagged with a $^\dagger$, a full optimization has not been
performed and possible improvements have not been studied yet.}
\label{tab:parameters}
\end{table}

The L.E. target is still made of graphite, but the spaces have been
eliminated. A more compact target ( 1 meter full length) allows for a better
focusing and increases the low-energy yield as will be discussed in the
next section.

The decay tunnel length in the present calculations has been set at 350
meters. This corresponds to about two decays lengths for pions producing
2.5 GeV neutrinos.
   
The 
neutrino energy of interest correspond to pions in the  range 0.7-5.5 GeV. 
To focus
these pions, we adopt a standard double-horn system (following the CNGS 
tradition, we call {\it horn} the first magnetic lens and {\it reflector}
the second). Both have to be 
placed near to or even around 
the target, to capture particles emitted at relatively large angles.
The average transverse momentum of secondary particles is around 300 MeV/c,
with a sizable number of events up to 600 MeV/c, corresponding to 750 mrad
for 1 GeV/c pions. These particles have to be bent before they travel too far 
away in radius, therefore the horn magnetic field has to be high
enough.  This also means that the horn should be shorter than the ones used
to focus high energy beams, because the particles should not travel 
in the magnetic field for a distance longer than their curvature radius. 

We obtained good focusing capability with two four meter long horns. The
horn current has been set at 300kA, the reflector one at 150kA. The horn
starts 25~cm after the target entrance face, the reflector starts just two
meters after the horn end. Horn
and reflector shapes has been computed to focus 2~GeV/c and 3~GeV/c 
particles respectively. We are aware that these (parabolic)
 horn shapes are derived in the
approximation of point-like source, that is not verified in the present
case. However, the Monte Carlo calculations verified the good focusing
capabilities of this system. The focusing efficiency in the range of
interest is around 50\%. 

\section{Neutrino fluxes and rates}

With the standard CNGS parameters, the low-energy neutrino flux is low, as can
be seen from the entries flagged by $^\dagger$ in Table~\ref{tab:CNGS}, 
 even assuming perfect focusing: we expect 0.9 $\nu_\mu$ CC events per kton
per $10^{19}$ pots for the tau focusing. Even if ideal focusing is assumed, the rate
is only improved by a factor 2.

The more compact target, 
and a wider acceptance, give about a factor five higher rate in the 0.-2.5 GeV
range, or about 4.5 $\nu_\mu$ CC events per kton
per $10^{19}$ pots for the real focusing and up to 9.0 $\nu_\mu$ CC events per kton
per $10^{19}$ pots for the ideal focusing.

\tabcolsep=.8mm
\begin{table}[tb]
\begin{tabular}{|c|c|c|c|c|c|c|c|c|c|}
\hline
&&&&&\multicolumn{2}{c|}{ $10^{19}$ p.o.t. 
}&\multicolumn{2}{c|}{$<E_\nu>$, CC}&\\
$E_p$&focus&decay tunnel &$\nu_\mu$ flux & $\nu_e$ flux&$\nu_\mu$ CC & $\nu_e$ CC 
&$\nu_\mu$ &$\nu_ e$ & $\nu_e / \nu_\mu$ \\  
GeV&&length (m)&\multicolumn{2}{c|}{$\nu$/cm$^2$}&
\multicolumn{2}{c|}{ev/kton}&
\multicolumn{2}{c|}{GeV }&CC\\
& & & & & & & & &\\
\hline
400 & p.f  &350
	&$ 1.3\cdot 10^{-13}$
	&$ 2.6\cdot 10^{-15}$
	&$ {\bf 9.0}$
	&$ 0.12$
&1.8&1.8&1.3\% \\
400 & horn &350
	&$ 1.0\cdot 10^{-15}$
	&$ 9.0\cdot 10^{-16}$
	&$ {\bf 4.5}$
	&$ 4.2\cdot 10^{-2}$
&1.8&1.4&0.9\% \\
400 & p.f $^\dagger$&CNGS
	&$ 1.6\cdot 10^{-14}$
	&$ 3.2\cdot 10^{-16}$
	&$ {\bf 1.8}$
	&$ 2.2\cdot 10^{-2}$
&2.1&1.7&1.2\% \\
400 & $\tau ^\dagger$&CNGS
	&$ 1\cdot 10^{-14}$
	&$ 9.4\cdot 10^{-17}$
	&$ {\bf 0.9}$
	&$ 8.7\cdot 10^{-3}$
&1.8&1.8&0.9\%\\

\hline
\hline
\end{tabular}
\caption{Neutrino beam parameters for the 
CNGS baseline, with $E_\nu <2.5$~GeV. The $^\dagger$ cases correspond to
the {\it present CNGS design} for target, acceptance and focusing system.}
\label{tab:CNGS}
\end{table}

\begin{figure}[p]
\centering
\epsfig{file=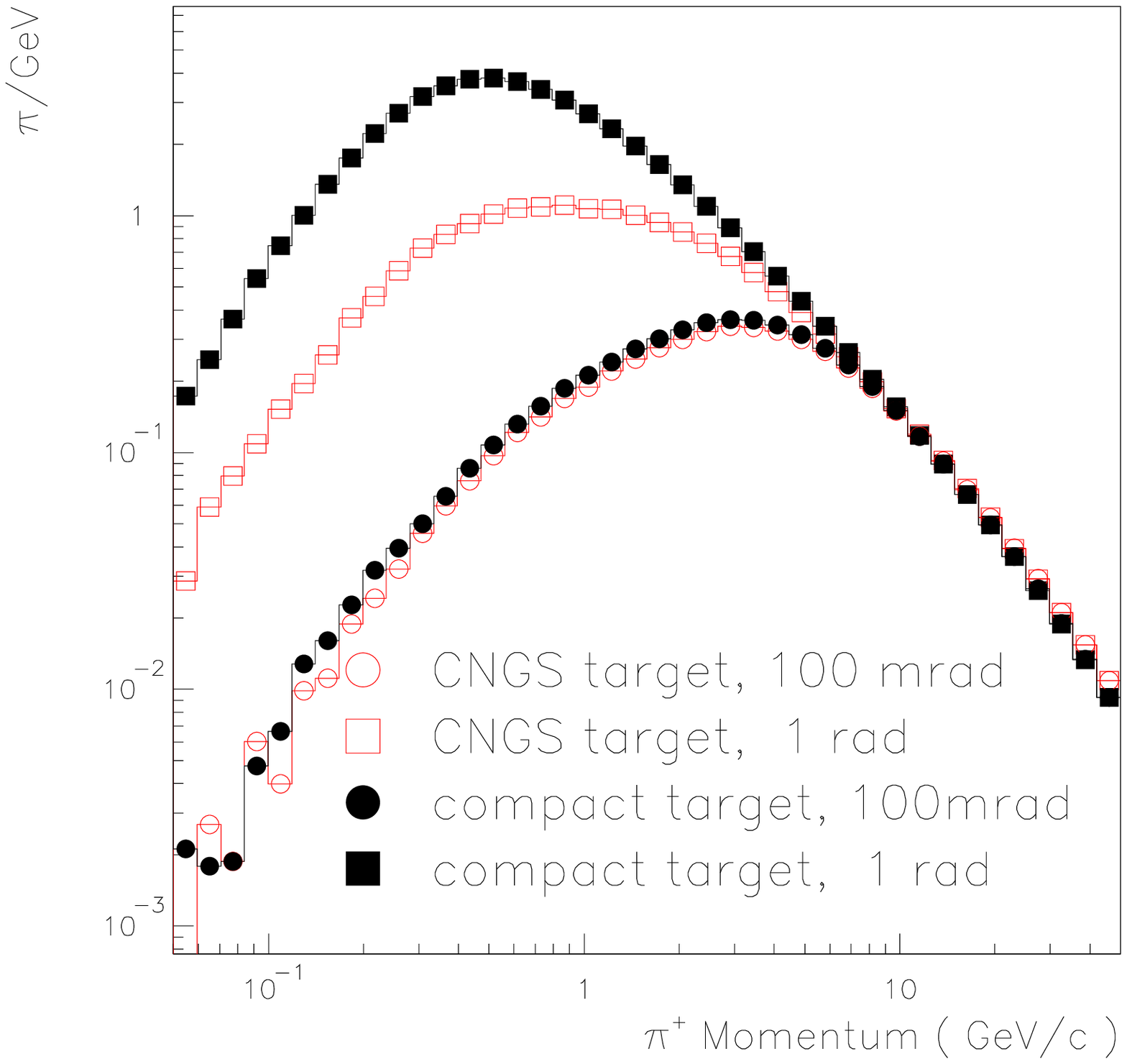,width=18cm}
\caption{Pion production at 400 GeV, for the CNGS target within 100 mrad
  acceptance, for a compact target ( the one of this study) within 100mrad
  acceptance and within 1 rad acceptance}
\label{fig:prod400}
\end{figure}

While in both cases the focusing introduces an efficiency factor of 
about 50\% with respect to the ideal focusing, our improvement
comes from the ability to capture more wide angle, soft pions.
The interplay of target configuration and angular acceptance 
can be seen clearly from Figure
\ref{fig:prod400}: the low-energy part of the produced pion spectrum is
enhanced  due to the higher re-interaction probability in a compact target, 
at the expenses of the medium-high energy part, 
the one of interest for $\tau$ appearance.  This enhancement is small in
the forward direction (100 mrad acceptance), 
but it grows at larger angles (up to 1 rad), where it reaches an
average factor of 1.5 for pions in the 1-5 GeV energy range. The real boost
in low energy 
pion production comes however from the angular acceptance of the system, 
which accounts for a factor 3.3 in the case of a compact target and a
factor 2.3 in the case of the CNGS target.
The difference in acceptances 
(1 rad in the low-energy option vs. 100 mrad for CNGS in perfect focusing) 
is therefore the dominant factor.

\begin{figure}[p]
\centering
\epsfig{file=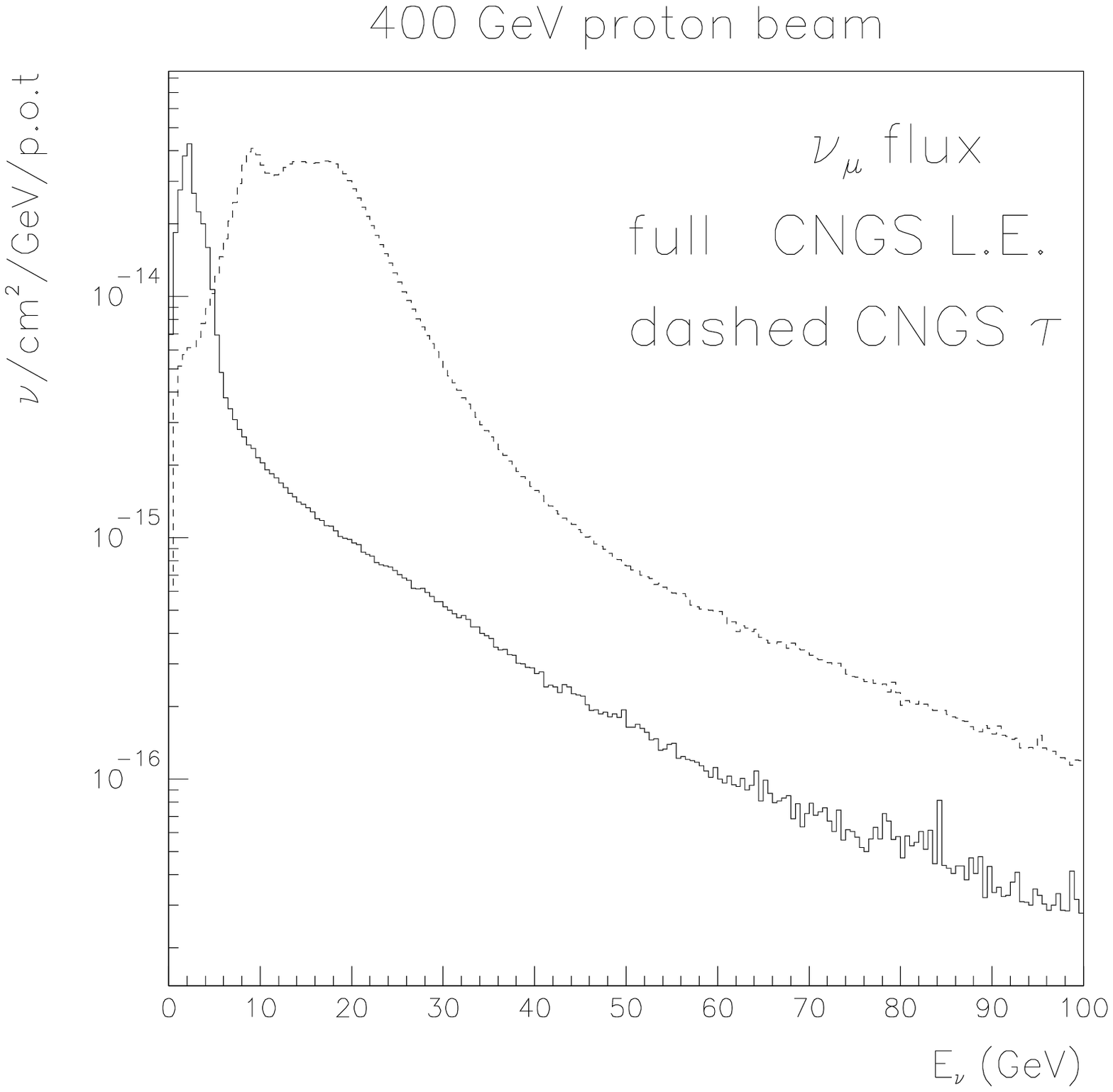,width=11cm%
,bbllx=0pt,bburx=500pt,bblly=40pt,bbury=520pt}
\epsfig{file=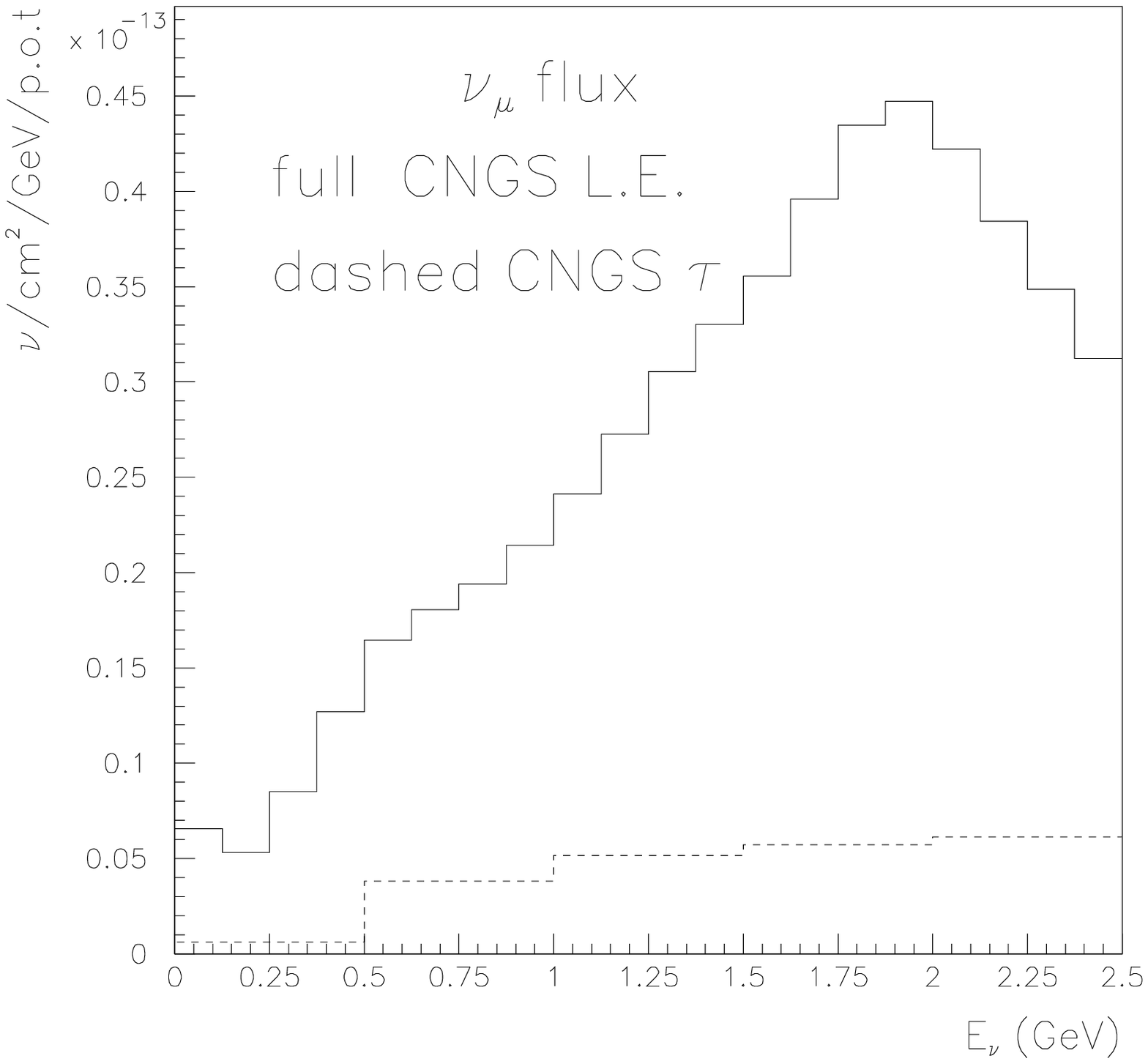,width=11cm
,bbllx=0pt,bburx=500pt,bblly=40pt,bbury=500pt}
\caption{ Muon neutrino fluxes at Gran Sasso, for the present CNGS
  design and for the new target/optics configuration. Whole spectra (top)
  and low-energy(bottom) are shown}
\label{fig:fluxmu}
\end{figure}
\begin{figure}[p]
\centering
\epsfig{file=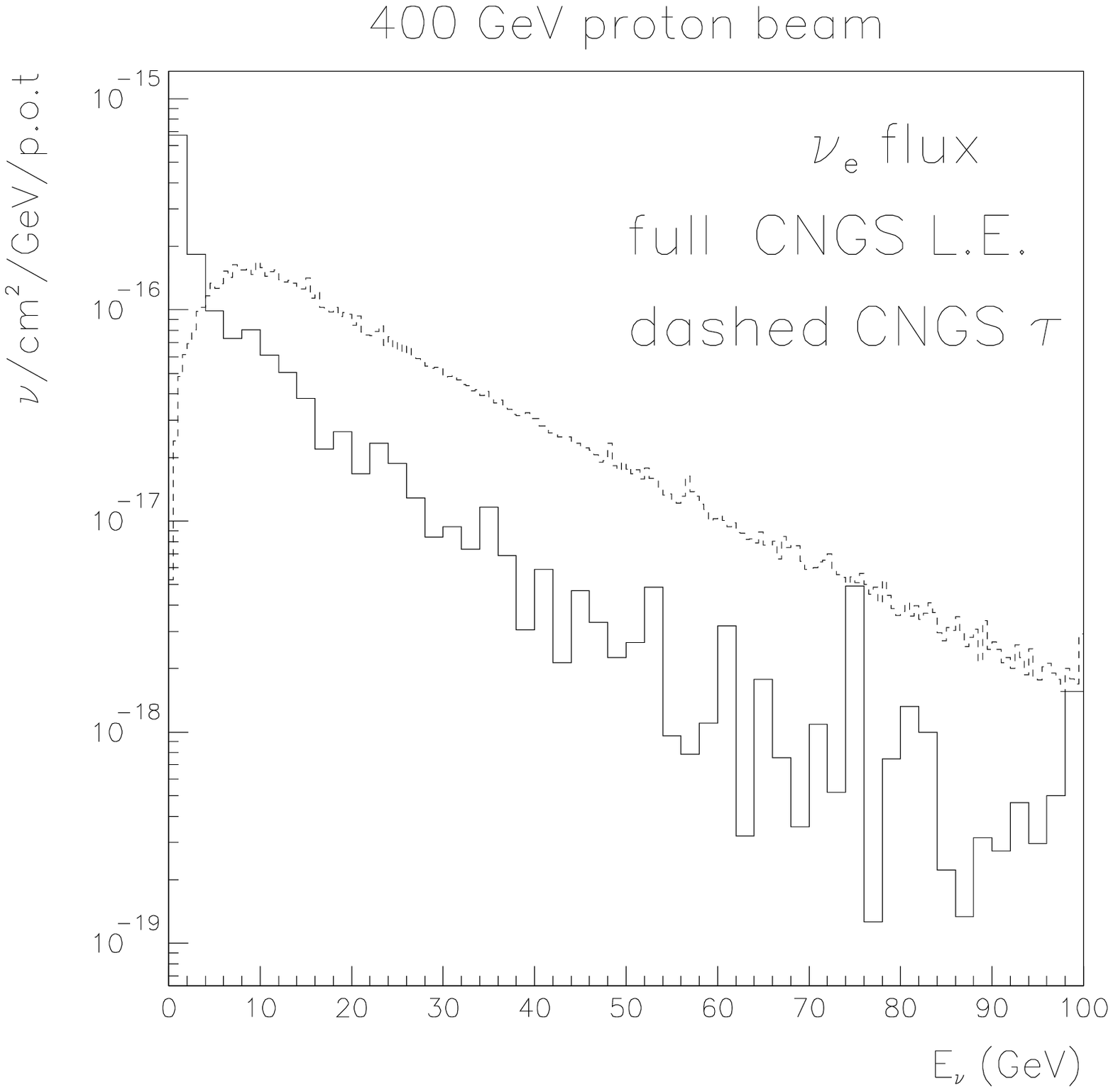,width=11cm
,bbllx=0pt,bburx=500pt,bblly=40pt,bbury=520pt}
\epsfig{file=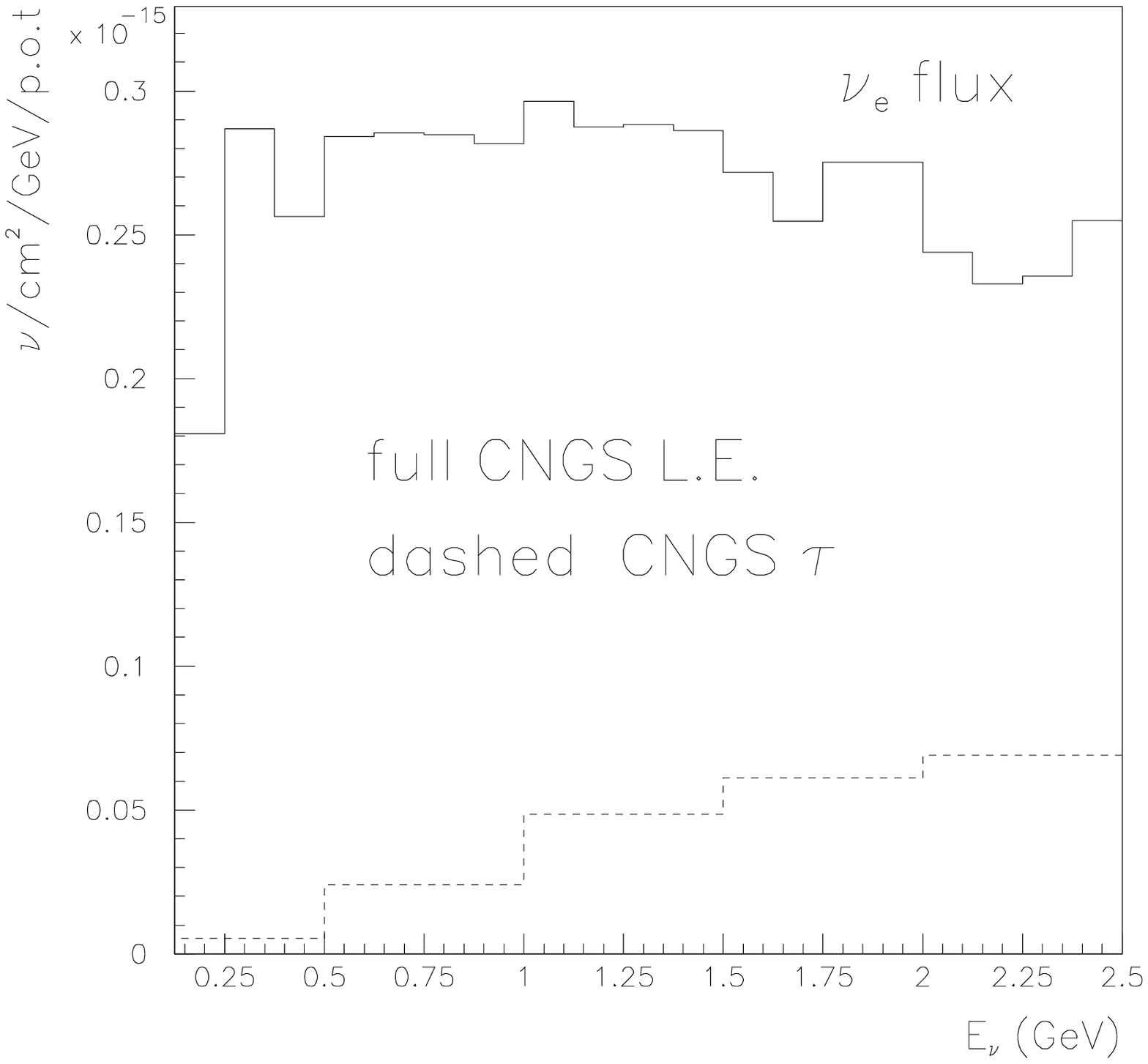,width=11cm
,bbllx=0pt,bburx=500pt,bblly=40pt,bbury=500pt}
\caption{ Electron neutrino fluxes at Gran Sasso, for the present CNGS
  design and for the new target/optics configuration. Whole spectra (top)
  and low-energy(bottom) are shown}
\label{fig:fluxe}
\end{figure}

Even after focusing, the L.E. option produces five times more low energy
$\nu_\mu$ than the standard one. The difference can be appreciated by eye
in Figure~\ref{fig:fluxmu}, where we show the $\nu_\mu$ fluxes at Gran-Sasso 
for the two configurations. The electron neutrino contamination is of the
same order of magnitude (around 1\%) for the $\tau$ and
L.E. options. Electron neutrino spectra are plotted in Figure~\ref{fig:fluxe}.


\section{A near detector ?}
During the optimization for low energy pions, it came natural
to reduce the ``used'' length of the decay tunnel from the 
available 1~km of the CNGS, in order
to reduce the high energy neutrino tail. 

The reduction of the actual effective decay length could
be accomplished by relocating some of the graphite and iron
blocks of the currently planned CNGS beam dump within the
decay tunnel. We are aware that this operation
is technically not as trivial as it might sound, however, we
want to point out that it is not necessary to fill the entire
decay tunnel but simply to stop the hadrons at the given point
of the decay tunnel in order to achieve our wanted result.

The reduced use of the decay tunnel leaves the room for an eventual
``near'' detector to be placed in the present beam dump position, that is
at about 1 km from the target. This would allow us to monitor
the beam in absence of oscillations and to predict the beam
spectrum at the far position.

\begin{figure}[p]
\centering
\epsfig{file=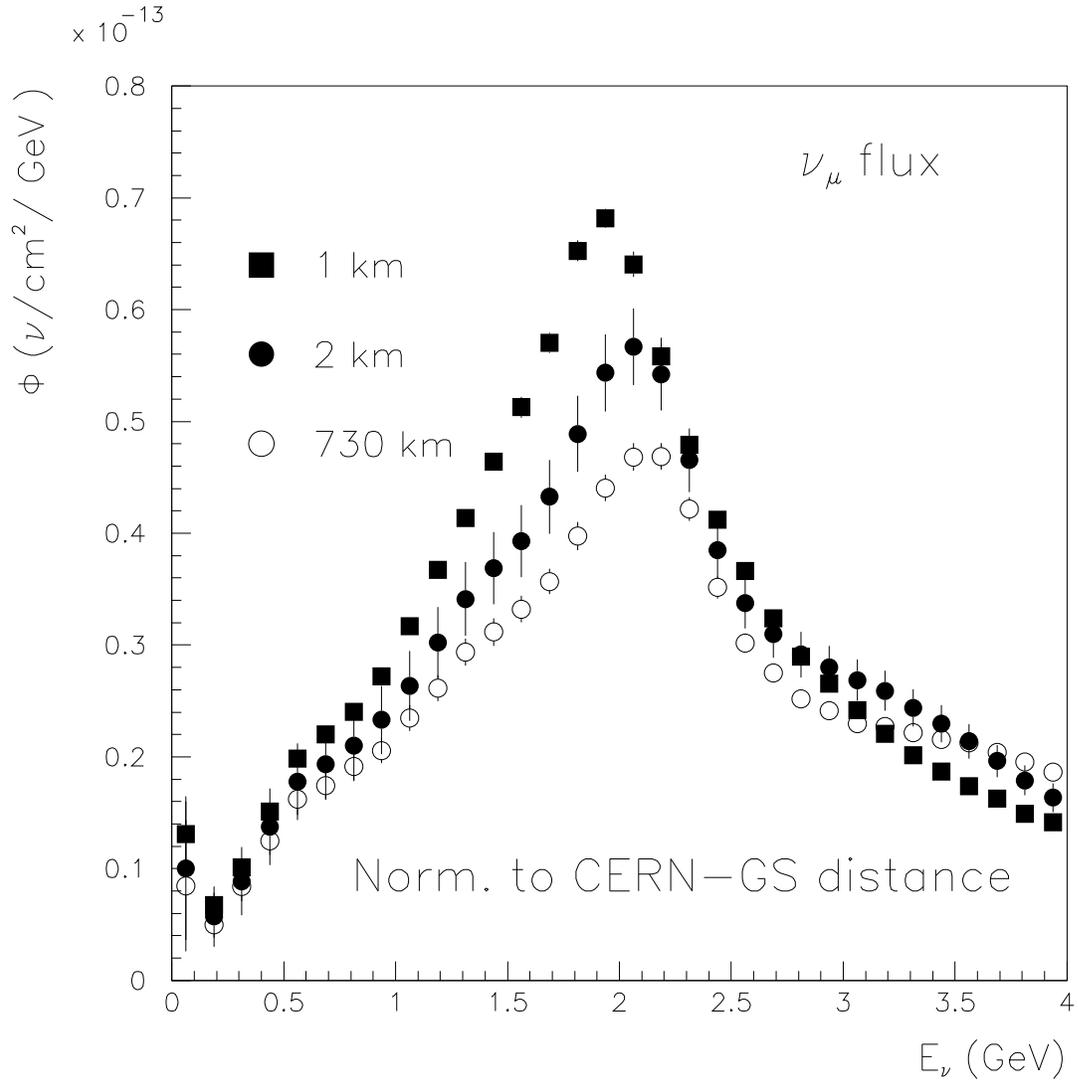,width=17cm}
\caption{ Muon neutrino fluxes at Gran Sasso and at ``near''
  positions, one and two kilometers from the target.
 For direct comparisons, the flux for the 
near detectors have been rescaled according to the square of 
 (target-near)/(target-Gran Sasso) distances.}
\label{fig:far-near}
\end{figure}

\begin{figure}[p]
\centering
\epsfig{file=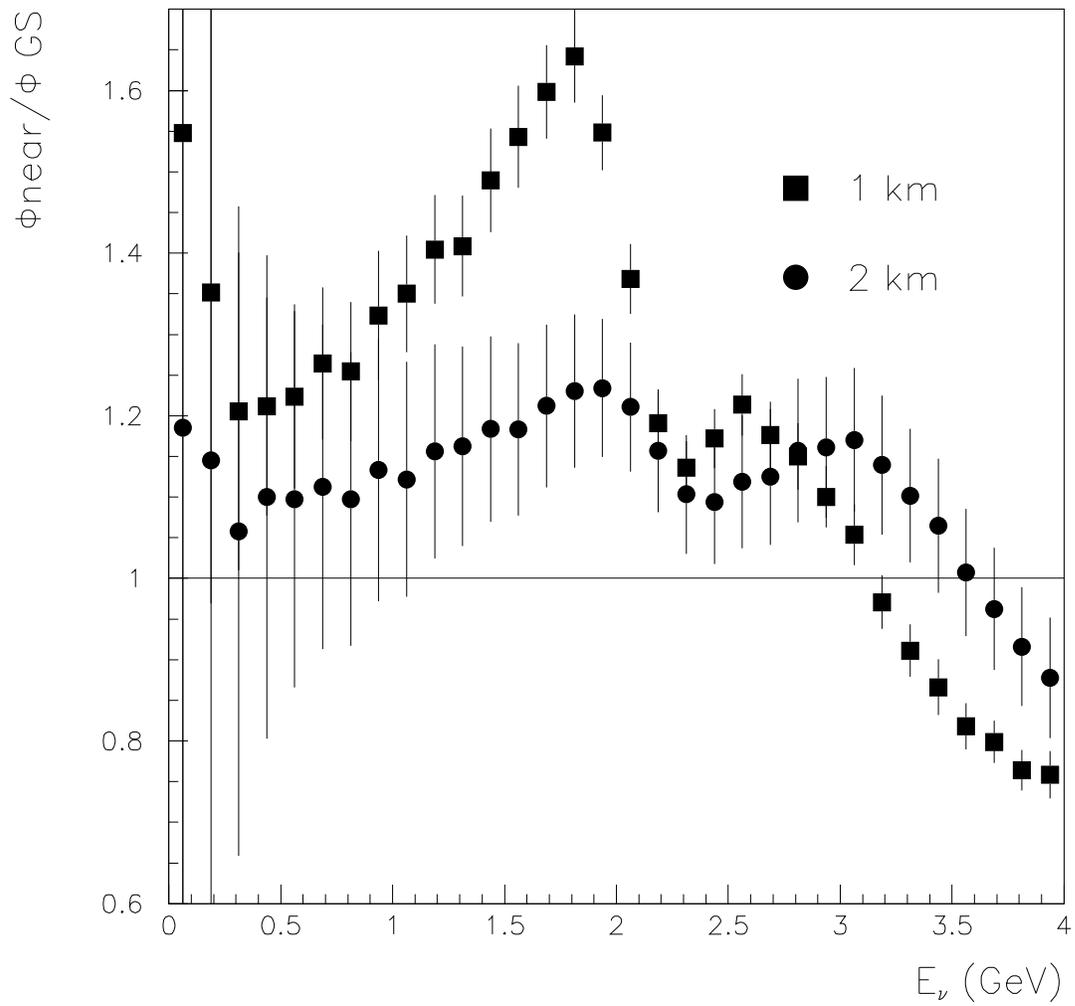,width=17cm}
\caption{ Ratio of near/far muon neutrino fluxes. }
\label{fig:far-near-ratio}
\end{figure}

As is well known however\cite{minos,k2k,JHF}, the
neutrino spectra at the near position are different from the ``far'' ones,
mainly because they feel the finite size of the decay tunnel. 
We have simulated for the L.E.  beam the $\nu_\mu$ fluxes
as a function of $\nu$ energy at 1 and 2 km from the target. 
The 1~km distance would correspond to the current ``beam dump area''
while the 2~km corresponds to the originally planned location of
the TOSCA detector. We have rescaled our results according 
to the square of the baseline ratio, such that the spectra are
directly comparable.  As shown in 
Figures~\ref{fig:far-near} and \ref{fig:far-near-ratio}, the low
energy 
part of the spectrum is enhanced with respect to the one at Gran Sasso,
with differences up to 60\% in the case of the very near
detector.

Therefore, the of $\nu_e$ beam contamination at Gran Sasso cannot be 
directly evaluated from the measurement at the near detector, but has to be
``propagated'' through Monte Carlo simulations. 
Hence, the need of the near detector is not fundamental, however, it would
allow to cross-check the beam simulation in the region of no oscillations
which can then be extrapolated at far distance. The resulting systematic
error has not been estimated yet, however, we point out in next section
that our results will be limited by statistics.

\begin{table}[tbh]
\begin{center}
 \begin{tabular}{|c|c|c|c|c|c|}
\hline
$\theta_{13}$&sin$^2 2\theta_{13}$&\multicolumn{2}{c|}{$\nu_e$ CC} &
\multicolumn{2}{c|}{$\nu_\mu \rightarrow \nu_e$ }\\
(degrees)&&$E_\nu<4$ GeV&$E_\nu < 50$ GeV&$E_\nu<4$ GeV&$E_\nu < 50$ GeV\\
\hline
  9&  0.095& 5& 44 &  16&   22.\\
  8&  0.076&  5& 44 &  13&   18.\\
  7&  0.059& 5& 44 &    10&   13\\
  5&  0.030& 5& 44 &    5&    7\\
  3&  0.011&  5& 44 &   1.8&    2.5\\
  2&  0.005&  5& 44 &   0.8&    1.1\\
  1&  0.001&  5& 44 &   0.2&    0.3\\
\hline
\end{tabular}
\end{center}
\caption{Events from the CNGS L.E. beam, 
assuming $\Delta$m$^2_{23}=3 \times 10^{-3} {\mathrm eV}^2$, 
 $\theta_{23}=45^\circ$, 5 years of operation and 2.35 kton fiducial mass.} 
\label{tab:evtle}
\end{table}

\begin{table}[tbh]
\begin{center}
\begin{tabular}{|c|c|c|c|c|c|}
\hline
$\theta_{13}$&sin$^2 2\theta_{13}$&\multicolumn{2}{c|}{$\nu_e$ CC} &
\multicolumn{2}{c|}{$\nu_\mu \rightarrow \nu_e$ }\\
(degrees)&&$E_\nu<4$ GeV&$E_\nu < 50$ GeV&$E_\nu<4$ GeV&$E_\nu < 50$ GeV\\
\hline
9&  0.095& 1.5 & 150&    4&   42\\
  8&  0.076&   1.5 & 150&  3.1&   34\\
  7&  0.059&   1.5 & 150&  2.4&   26\\
  5&  0.030&    1.5 & 150& 1.2&   14\\
  3&  0.011&    1.5 & 150& 0.4&    5\\
  2&  0.005&    1.5 & 150& 0.2&    2.2\\
  1&  0.001&    1.5 & 150& 0.1&    0.5\\
\hline
\end{tabular}
\end{center}
\caption{Events  from the CNGS $\tau$ beam, 
assuming $\Delta$m$^2_{23}=3 \times 10^{-3} {\mathrm eV}^2$, 
 $\theta_{23}=45^\circ$, 5 years of operation and 2.35 kton fiducial mass.} 
\label{tab:evtcngs}
\end{table}

\section{Oscillated events and limits}
\begin{figure}[p]
\centering
\epsfig{file=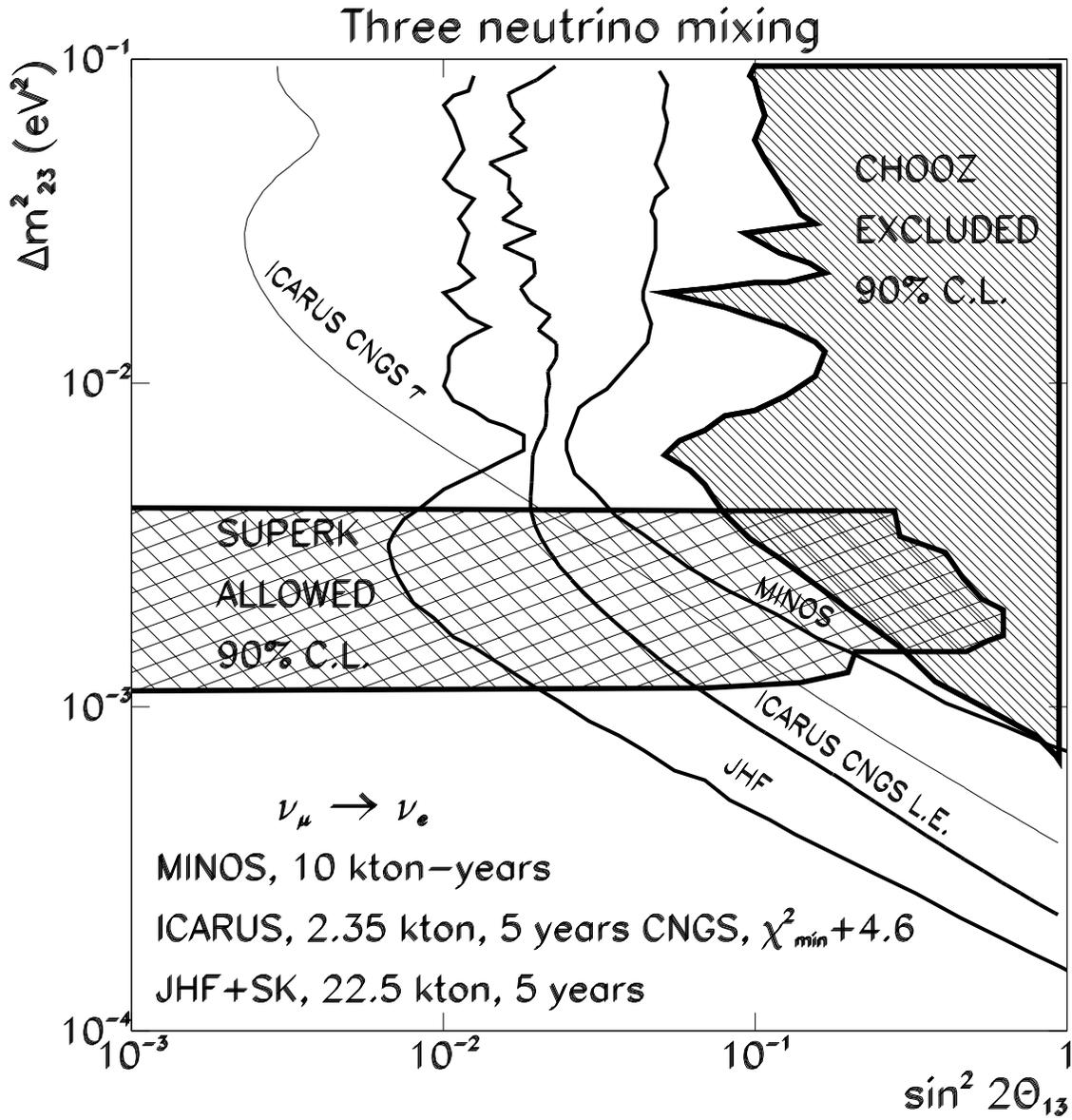,width=17cm}
\caption{Expected sensitivity 
to $\sin^22\theta_{13}$ mixing angle at nominal CNGS intensity, 
compared to existing results from CHOOZ\cite{Chooz} and SuperK\cite{nu2000} and expected results from
MINOS\cite{minos} and JHF-SK\cite{JHF}.}
\label{fig:exclus}
\end{figure}

Figure~\ref{fig:exclus} shows the exclusion plots for $\sin^22\theta_{13}$
at the 90\% C.L. for the CNGS $\tau$ and our optimized CNGS L.E. options, 
compared with the CHOOZ limit, the SuperK allowed region and 
other proposed experiments.
These contours have been derived from $\nu_e$ appearance, 
assuming 5 years of operation at the
nominal CNGS intensity, and an ICARUS
fiducial volume of 2.35~kton. For MINOS we assume an exposure
of 10~$\rm kton\times year$\cite{nu2000minos} and we quote
the JHF limit according to the proposal of the OAB beam\cite{JHF}.
Three neutrino formalism and maximal $\numu$-$\nutau$ mixing have been assumed, 
i.e. $\theta_{23}=45^o$, so that the usual ``appearance
factor'' $P(\nu_\mu\ra\nu_e)=0.5\times P(\nu_e\ra\nu_\mu/\nu_\tau)$
has been taken into account.

Only the intrinsic
$\nue$ beam contamination has been  considered here for the
background evaluation;
other background sources, such as $\pi^0$ production in neutral current events,
have been extensively studied in the past\cite{icarus} and found to be negligible
given the excellent granularity of the ICARUS detector. They will nonetheless
be precisely re-addressed in a forthcoming paper~\cite{Ferrari:2002yj}. We however
do not expect any significant changes in sensitivity introduced by other sources of backgrounds.

\begin{figure}[p]
\centering
\epsfig{file=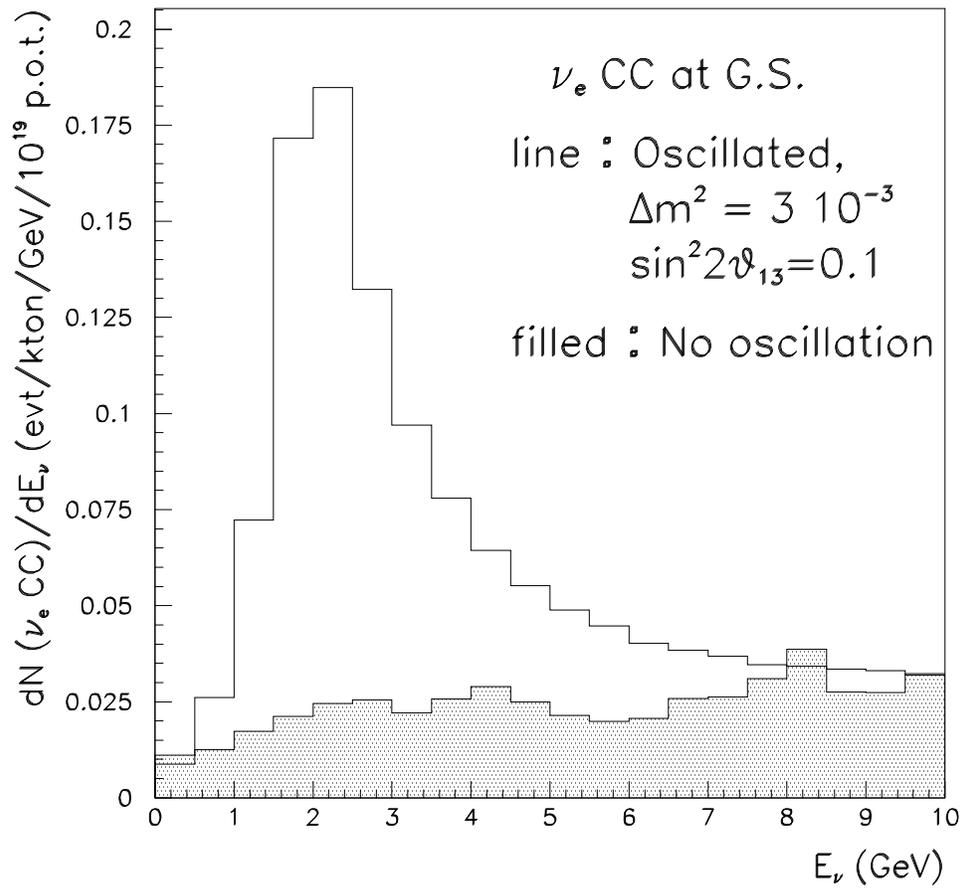,width=15cm}
\caption{Comparison of 
$\nue$ CC spectra at Gran Sasso, in presence and in absence of oscillation.}
\label{fig:eoscicomp}
\end{figure}

The integrated number of background and
oscillated $\nue$ events in the hypothesis 
$\Delta m^2 = 3 \times 10^{-3} \mathrm{eV}^2$ are listed in
Table~\ref{tab:evtle}. The oscillation maximum in the case of 
$\Delta m^2 = 3 \times 10^{-3} \mathrm{eV}^2 , \sin^22\theta_{13}=0.1 $ is
evident from the event spectra shown in Figure~\ref{fig:eoscicomp}.
For comparison, we also quote the numbers for the CNGS $\tau$
optimization in Table~\ref{tab:evtcngs}.

The CNGS L.E. options represents an increase of roughly a factor 2 in
sensitivity with respect to the CNGS $\tau$ beam, 
within the SuperK allowed region. For $\Delta m^2=2.5\times 10^{-3}\rm\ eV^2$,
we find:
\begin{eqnarray}
(\sin^22\theta_{13})_{CNGS,\tau}< 0.04 \rm\ \ \ or\ \ \  \theta_{13}<6^o \\
(\sin^22\theta_{13})_{CNGS,L.E.}< 0.02\rm\ \ \ or\ \ \  \theta_{13}<4^o
\end{eqnarray}
 The improvement over the 
CHOOZ limit is almost tenfold\cite{Chooz}, and the performances are three times
better than those foreseen by MINOS\cite{nu2000minos}:
\begin{eqnarray}
(\sin^22\theta_{13})_{CHOOZ}< 0.14 \rm\ \ \ or\ \ \  \theta_{13}<11^o\\
(\sin^22\theta_{13})_{MINOS}< 0.06\rm\ \ \ or\ \ \  \theta_{13}<7^o
\end{eqnarray}

Finally, the JHF proposal with the OAB beam gives\cite{JHF}
\begin{eqnarray}
(\sin^22\theta_{13})_{JHF,OAB}< 0.006 \rm\ \ \ or\ \ \  \theta_{13}<2.2^o
\end{eqnarray}
The comparison with the first phase (5 years) of the superbeam JHF is still 
slightly unfavourable, however we point out
the possible time schedule and the probable SPS  proton 
beam intensity upgrades.

\begin{itemize}
\item The JHF-Kamioka experiment will start one or two years later than the CNGS
beam\footnote{We
assume the starting dates of 2007 for CNGS and 2009 for JHF.}; 
this allows to set at 7 years the CNGS L.E. data taking. 

\item As for proton intensity, it is expected that PS and SPS upgrades\cite{cappi}
will bring the
accelerated intensity from $4.8\times 10^{13} $ to $7\times 10^{13} $
protons per cycle. This represents an increase of $\approx 50\%$ in the p.o.t. per year.
This intensity upgrade is already taken into account in all the
design specifications for the CNGS facility\cite{CNGSNW}. We call
this CNGS1.5.
\end{itemize}

\begin{figure}[p]
\centering
\epsfig{file=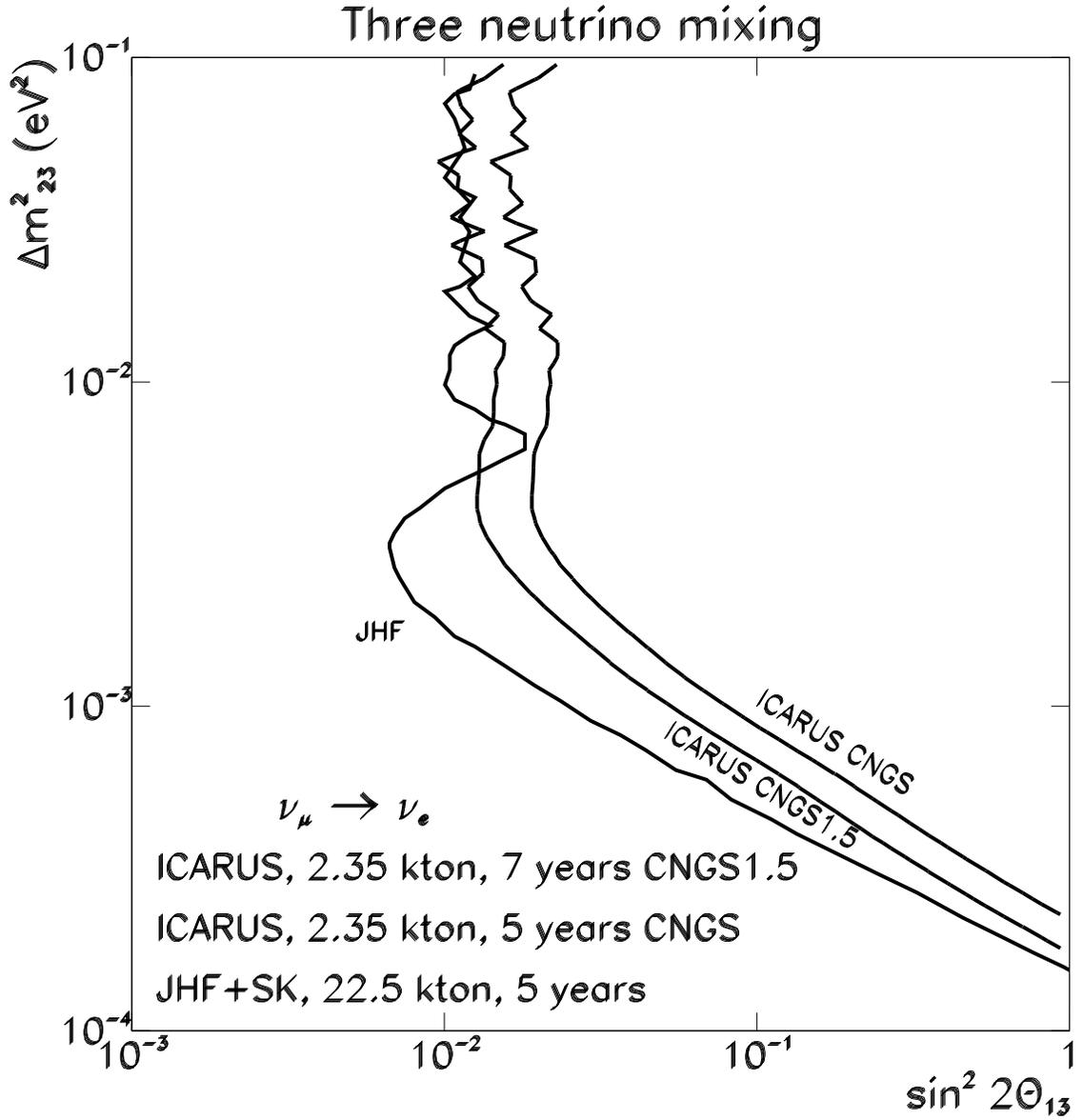,width=17cm}
\caption{Comparison of expected sensitivity to $\sin^22\theta_{13}$ mixing angle with
an improved CNGS$\times 1.5$\cite{cappi} and 7 years of running.}
\label{fig:exclustwo}
\end{figure}

Accordingly, the sensitivity contour for 7 years data taking at CNGS1.5 nominal
intensity is shown in Figure~\ref{fig:exclustwo}. This gives for 
$\Delta m^2=2.5\times 10^{-3}\rm\ eV^2$
\begin{eqnarray}
(\sin^22\theta_{13})_{CNGS1.5,L.E.}< 0.015 \rm\ \ \ or\ \ \  \theta_{13}<3.5^o
\end{eqnarray}

Since we are limited by the intrinsic $\nu_e$ background, the sensitivity
scales with the square of the exposure and one would need
to increase the intensity of the CNGS and/or the total mass of ICARUS by 
a factor
\begin{eqnarray}
\left(\frac{(\sin^22\theta_{13})_{CNGS1.5,L.E.}}{(\sin^22\theta_{13})_{JHF,OAB}}\right)^2
\approx 6
\end{eqnarray}
in order to reach exactly the level of sensivity of the JHF superbeam.

\section{Conclusions}

The CERN-NGS has been originally optimized and coupled to high quality
detectors in order to unambiguously give evidence for the $\nu_\mu\ra\nu_\tau$
flavor oscillation mechanism.

However given the CERN financial situation, the CERN-NGS programme, approved in 1999,
is now foreseen to start with full intensity in the year 2007.
Given these delays, it is worth wondering if the priority
of the program should be tau appearance (``a confirmation of the oscillation
mechanism'') or electron appearance (``a potential discovery of a new flavor
transition'').

In this paper, we have studied a different optimization of the CNGS beam optics
that is optimized for $\nu_\mu\ra\nu_e$ oscillation searches at the 
$\Delta m^2$ indicated by the atmospheric neutrinos. We find that this beam
coupled with the approved ICARUS T3000 experiment at LNGS would offer great
opportunities to find neutrino oscillations driven by $\theta_{13}$ on
the same time scale as that of the proposed JHF superbeam.

To reach a sensivity $\sin^22\theta_{13}<0.006$, the total exposure
should be increased by a factor 6 compared to what we have assumed here.
This would require a big increase in beam intensity or a substantial increase
of the liquid argon mass. The cost of multiplying the mass of ICARUS by a factor six
is interestingly on the scale of the price of a new superbeam.

\section*{Acknowledgments}
We thank A.~Bettini for fruitful discussions on the $\theta_{13}$
sensitivity of ICARUS at LNGS. We acknowledge A.~Ferrari for his
invaluable help in the 
development of the FLUKA-CNGS simulation code without
which this work would have been impossible.

    
%
%



\end{document}